\documentclass[aip,cha,numerical,reprint,eqsecnum,floats]{revtex4-1} 
	
\usepackage[english]{babel}
\usepackage{amsmath}
\usepackage{amssymb}
\usepackage{mathrsfs}
\usepackage{amsthm}
\usepackage{textcomp}
\usepackage{xcolor}
\usepackage{graphicx}
\usepackage{enumerate}
\usepackage{mathtools}
\usepackage{hyperref}
\usepackage[title]{appendix}
\usepackage{bm}

\newcommand{\LL}{\mathbb{L}}

\begin{document}

\title{Rate of change of frequency under line contingencies in high voltage electric power networks with uncertainties}

\author{Robin Delabays}
\email{robin.delabays@hevs.ch}
\affiliation{School of Engineering, University of Applied Sciences of Western Switzerland, CH-1950 Sion, Switzerland}
\affiliation{Automatic Control Laboratory, Swiss Federal Institute of Technology (ETH), CH-8092 Z\"urich,  Switzerland}

\author{Melvyn Tyloo}
\affiliation{School of Engineering, University of Applied Sciences of Western Switzerland, CH-1950 Sion, Switzerland}
\affiliation{Institute of Physics, \'Ecole Polytechnique F\'ed\'erale de Lausanne (EPFL), CH-1015 Lausanne, Switzerland.}

\author{Philippe Jacquod}
\affiliation{School of Engineering, University of Applied Sciences of Western Switzerland, CH-1950 Sion, Switzerland}
\affiliation{Department of Quantum Matter Physics, University of Geneva, CH-1211 Geneva, Switzerland}

\date{\today}

\begin{abstract}
 In modern electric power networks with fast evolving operational conditions, assessing the impact of contingencies is becoming more and more crucial. 
 Contingencies of interest can be roughly classified into nodal power disturbances and line faults. 
 Despite their higher relevance, line contingencies have been significantly less investigated analytically than nodal disturbances. 
 The main reason for this is that nodal power disturbances are additive perturbations, while line contingencies are multiplicative perturbations, which modify the interaction graph of the network. 
 They are therefore significantly more challenging to tackle analytically. 
 Here, we assess the direct impact of a line loss by means of the maximal Rate of Change of Frequency (RoCoF) incurred by the system. 
 We show that the RoCoF depends on the initial power flow on the removed line and on the inertia of the bus where it is measured. 
 We further derive analytical expressions for the expectation and variance of the maximal RoCoF, in terms of the expectations and variances of the power profile in the case of power systems with power uncertainties. 
 This gives analytical tools to identify the most critical lines in an electric power grid.
\end{abstract}

\maketitle

\begin{quotation}
 Electrical networks can be subjected to many types of disturbances, such as small to medium  fluctuations of the power injections and consumptions due to intermittent energy sources, larger power disturbances such as power plant outages, or the breakdown of an electrical line. 
 There are various measures of the impact of such disturbances. 
For instance, one may consider the time the system requires to settle back to a synchronous state, or the magnitude of the excursion of some quantities, such as voltage angles or frequency, away from their desired values. 
 In this work, we  investigate the effect of a line loss on the Rate of Change of Frequency (RoCoF), which is the largest time derivative of the voltage frequencies. 
 It is a measure of the severity of a perturbation that is standardly used in the operation of electric power grids. 
 In practice, the current operating state of the system is not known. 
 We therefore derive statistical properties of the RoCoF after a line loss for cases when the exact operating state of the system is uncertain. 
\end{quotation}

\section{Introduction}\label{sec:intro}
Modern societies are heavily  relying on electric power. 
The ongoing energy transition in many parts of the world, aiming at substituting conventional power plants by new renewable energy sources is leading to significant changes in the operation of power systems. 
In this context, the reliability of electric power grids and the safety of electricity supply are becoming more challenging than they were in the past. 
Fast, preferably real-time assessments of network vulnerabilities are needed to guarantee a secure electricity supply. 
One appealing way to evaluate the robustness and identify vulnerabilities of electric power networks is to measure their dynamical response to various external perturbations. 
Different performance metrics have been introduced and used to qualitatively and quantitatively measure how strongly a given perturbation affects the system.

Commonly studied perturbations are changes in the power profile.~\cite{Sum15,Sia16,Far14,Tyl18a,Teg15,Gru18,Poo17,Pag17} 
They can reflect planned redispatches of power generation or uncontrollable variations of power consumptions. 
They may also result from more dramatic events, such as the unexpected loss of a power plant. 
Other important types of operationally relevant perturbations are line faults.~\cite{Col19}
As a matter of fact, electrical lines are vulnerable to various unpredictable events, such as grounding, overload, or attacks. 
Lines can also be voluntarily disconnected when the power flow they carry exceeds their thermal limit for too long, or when other safety limits are exceeded. 
Treating  line contingencies analytically is however much more challenging. 
Indeed, power perturbations appear in the dynamics as an additive disturbance, changing the power input or output at some buses, while a line loss modifies the network Laplacian matrix and appears as a multiplicative disturbance. 
Analytically evaluating the impact of a line contingency is then intrinsically more challenging than evaluating the impact of a power disturbance. 

In this paper, we are interested in the immediate impact of a line loss which is considered \emph{permanent}: 
the electrical line is disconnected and is not reconnected within a finite amount of time larger than the typical dynamical time scales in the system. 
In the high voltage power grids we are interested in, these other time scales are shorter than few seconds. 
When such a line loss is shorter than, say few minutes, it is reasonable to consider the power injections and consumptions as constant during the line fault. 
We focus on line losses that do not fragment the network as the loss of such lines usually lead to desynchronization. 
Such a disturbance requires a redispatch of the power generation, beyond our interest in this manuscript.

\begin{figure}
 \centering
 \includegraphics[width=.45\textwidth]{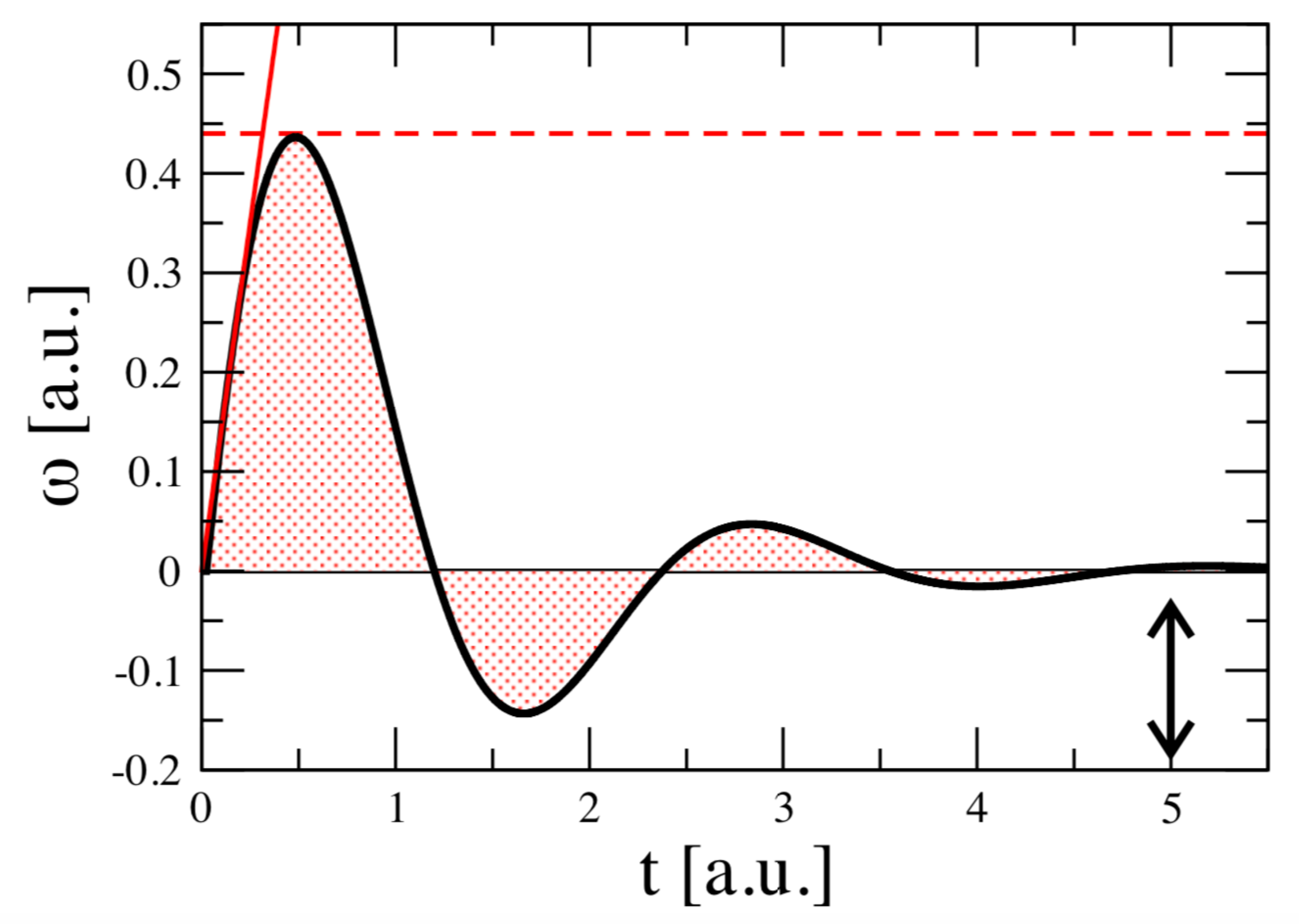}
 \caption{Sketch of the time-evolution of the voltage frequency following a fault at time $t=0$ (solid black line). 
 Different performance measures evaluate the magnitude of the fault, such as the frequency Nadir (dashed red line), the RoCoF (solid red line), and the recovery time (black arrow). 
 Performance measures based on ${\cal L}_2$-norms instead measure the overall magnitude of the excursion such as the squared red dashed area under the frequency curve.}
 \label{fig:pmeasures}
\end{figure}

The robustness of a dynamical system can be evaluated by its response to perturbations. 
Electric power quality is usually characterized by the amplitude of voltage and frequency variations. 
Here we neglect voltage angle variations and focus on the impact of disturbances on the frequency. 
Fig.~\ref{fig:pmeasures} illustrates the time evolution of the voltage frequency following a fault.
Starting from a stable steady state, the impact of a perturbation is to make the frequency oscillate
with an amplitude that is damped with time when things go according to plan. 
The strength of the perturbation is measured by the magnitude of some performance measure applied to the subsequent transient.~\cite{Bam12,Sum15,Sia16,Far14,Tyl18a}
Common performance measures include ${\cal L}_2$- and ${\cal L}_\infty$-norms of some system outputs -- based, e.g., on the voltage angle or frequency -- after a disturbance. 
In the scope of electrical transmission networks, ${\cal L}_2$-norms of the voltage angle deviation can be related to the transmission losses due to the transient~\cite{Teg15,Gru18} and ${\cal L}_2$-norms of the frequency deviations to the primary control effort following a disturbance.~\cite{Poo17,Col19} 
Standard measures of the impact of a disturbance commonly used by electrical engineers as decision variables can be formulated in terms of ${\cal L}_\infty$-norms.~\cite{Pag17} 
The maximal frequency deviation (Nadir) and the maximal RoCoF are respectively the ${\cal L}_\infty$-norm of the frequency deviation and of the time derivative of the frequency. 

In this work, we evaluate the impact of a disturbance by the RoCoF following it. 
Instead of measuring the global RoCoF, i.e., the maximal derivative of the mean frequency of the  network,~\cite{Pag17} we are interested in the \emph{maximal local RoCoF}, which we define as the maximal derivative of the frequency at any bus. 
The main reason for this choice is that the local change of frequency will affect security devices at the buses and might trigger disconnection of some elements, potentially leading to cascading events. 
Anticipating such events requires therefore to have a spatial resolution of frequency variations. 
From now on, we use the term RoCoF to refer to the maximal local RoCoF. 

Based on the linearized Swing Equations,~\cite{Ber00} we give an analytical estimate of the local RoCoF at both ends of the lost line, which is expected to be the largest RoCoF incurred by this disturbance. 
We see that the RoCoF directly depends on the power profile. 
In the case of unknown power profile, but with a knowledge of its probability distribution, we are able to determine the expectation and variance of the RoCoFs after a line contingency. 
Our results allow one to locate the most critical lines in a network.

\section{Model and approach}\label{sec:prelim}
To model line contingencies, we consider the dynamics of the linearized lossless Swing Equations after a line contingency. 
This is an often used approximation in very high voltage transmission networks, where angle differences are small in the operating state. 
The loss of a line changes the transmission network and our approach allows us to rely on the initial network Laplacian only.

\subsection{The model}\label{ssec:model}
We model the voltage phase dynamics by the linear lossless approximation of the Swing Equations~\cite{Ber00} 
\begin{align}\label{eq:swing}
 M\ddot{\bm{\theta}} + D\dot{\bm{\theta}} &= \bm{P} - \LL\bm{\theta}\, ,
\end{align}
where $\theta_i$, for $i\in\{1,...,n\}$, is the voltage phase angle at bus $i$, $M={\rm diag}(m_1,...,m_n)$ [$D={\rm diag}(d_1,...,d_n)$] is the diagonal matrix of inertias [dampings], $\bm{P}$ is the vector of powers injected ($P_i>0$) and consumed ($P_i<0$), satisfying $\sum_iP_i=0$, and $\LL$ is the weighted Laplacian matrix of the network's graph, with elements 
\begin{align}
 \LL_{ij} &= \left\{
 \begin{array}{ll}
  -b_{ij}\, , & \text{if } i\neq j\, , \\
  \sum_k b_{ik}\, , & \text{if } i=j\, .
 \end{array}
 \right.
\end{align}
The network's lines are modelled as lossless with susceptance $b_{ij}\geq0$, taking value zero if buses $i$ and $j$ are not connected.
The power flow they carry is approximated as $b_{ij} (\theta_i-\theta_j)$, in terms of the voltage angles at the buses they connect. 
This is a good approximation for very high voltage power grids in standard operational states where 
(i) lines have a conductance that is smaller than their susceptance by a factor of ten or more and 
(ii) voltage angle differences between connected buses do not exceed 30$^\circ$.~\cite{Kun94}

We denote the eigenvectors and eigenvalues of the weighted Laplacian matrix as $\bm{u}^{(\alpha)}$ and $\lambda_\alpha$, for $\alpha=1,...,n$. 
The Laplacian matrix $\LL$ always has a vanishing eigenvalue $\lambda_1=0$ associated with the constant eigenvector $\bm{u}^{(1)}=n^{-1/2}(1,...,1)$. 
Provided that the network is connected (which we always assume), all other eigenvalues of $\LL$ are positive, $0<\lambda_2\leq...\leq\lambda_n$. 
The fixed point of Eq.~\eqref{eq:swing} is obtained as 
\begin{align}\label{eq:init}
 \bm{\theta}^* &= \LL^\dagger\bm{P}\, , & \dot{\bm{\theta}}^* &= \bm{0}\, ,
\end{align}
where $\LL^\dagger$ is the Moore-Penrose pseudoinverse of $\LL$ defined as $\LL^\dagger\coloneqq T\cdot{\rm diag}(0,\lambda_2^{-1},...,\lambda_n^{-1})\cdot T$, and $T$ is the square matrix whose $i^{\rm th}$ column contains the components of $\bm{u}^{(i)}$. 

In general, some nodes, usually loads, are modeled as inertialess ($m_i=0$).~\cite{Ber81} 
However, in this manuscript, we will consider positive inertia at each node. 
This makes sense as future power grids are expected to have much more distributed energy sources, which will contribute to add inertia (physical or virtual\cite{Poo17}) everywhere in the network. 
Furthermore, we will numerically confirm that our analytical approach, assuming intertia everywhere, gives reasonable predictions in this more general case of inertialess loads.

\subsection{Line loss}\label{ssec:loss}
When the line between buses $i$ and $j$ is cut, the Laplacian matrix $ \LL^*$ of the resulting grid can be written as 
\begin{align}\label{eq:lap*}
 \LL^* &= \LL - b_{ij}\bm{e}_{ij}\bm{e}_{ij}^\top\, ,
\end{align}
where $\bm{e}_{ij}$ is the vector with $i^{\rm th}$ (resp. $j^{\rm th}$) component equal to $+1$ (resp. $-1$), and zero otherwise. 
This formulation makes it clear that a line loss corresponds to a rank-1 modification of the Laplacian matrix.

\subsection{The RoCoF}\label{ssec:rocof}
The RoCoF is defined as the change of frequency of the system on a finite time interval, because in real power grids, frequencies are monitored at discrete time intervals. 
RoCofs are then evaluated as the frequency slope between two such measurements and correspond to a discrete second derivative of the voltage angle. 

The RoCoF is often measured for the frequency averaged over the whole network.~\cite{Pag17}
As pointed out at the end of Sec.~\ref{sec:intro}, we are interested here in a local resolution of the RoCoF. 
We therefore consider the maximal local RoCoF which we define as the maximal time derivative of the frequencies over all nodes in the network, 
\begin{align}
 {\rm RoCoF} &\coloneqq \max_i \|\dot{\omega}_i(t)\|_\infty\, , & t &> 0\, ,
\end{align}
with $t=0$ indicating the time of the fault.
The maximal local RoCoF is always larger or equal than the global RoCoF. 

\subsection{Numerics}\label{ssec:numerics}
To confirm the results presented below in Sec.~\ref{sec:rocof}, we compare them to numerical simulations of the swing dynamics given by Eq.~\eqref{eq:swing} on the IEEE 118-Bus test case where we assume positive inertia at every node. 
The line weights $b_{ij}$ are the susceptances of the lines and the power profile $\bm{P}$ is proportional to the nominal power, $\bm{P}^*$, given in the test case. 

The inertia and damping parameters are constructed following the procedure described in Ref.~\onlinecite{Pag19}, to which we add a small random term to make sure none of our results are artifact of the chosen set of parameters. 
We have
\begin{align}
 m_i &= \frac{2H_iP_i^*}{\omega_0}\, , & d_i &= \gamma_i m_i\, ,
\end{align}
where $\omega_0=2\pi\cdot 50{\rm s}^{-1}$ ($60{\rm s}^{-1}$) is the nominal frequency of the network, $H_i$ is the inertia constant of the rotating machine at bus $i$, corresponding to the time over which the rated power of the machine provides a work equivalent to its kinetic energy of rotation, and $\gamma_i$ represents the ratio between damping and inertia at node $i$. 
We took these parameters uniformly in the intervals $\gamma_i\in [0.35,0.7]{\rm s}^{-1}$ and $H_i\in [5.0,7.0]{\rm s}$ (see Appendix 2 of Ref.~\onlinecite{Pag19} and Table 3.2 of Ref.~\onlinecite{Kun94} for more details).

\section{RoCoF estimate}\label{sec:rocof}
The analytical approach we present next is, we think, quite straightforward mathematically. 
Following its presentation, we verify its predictions numerically. 

\subsection{Analytical prediction}\label{ssec:analytic}
\begin{figure}
 \centering
 \includegraphics[width=.48\textwidth]{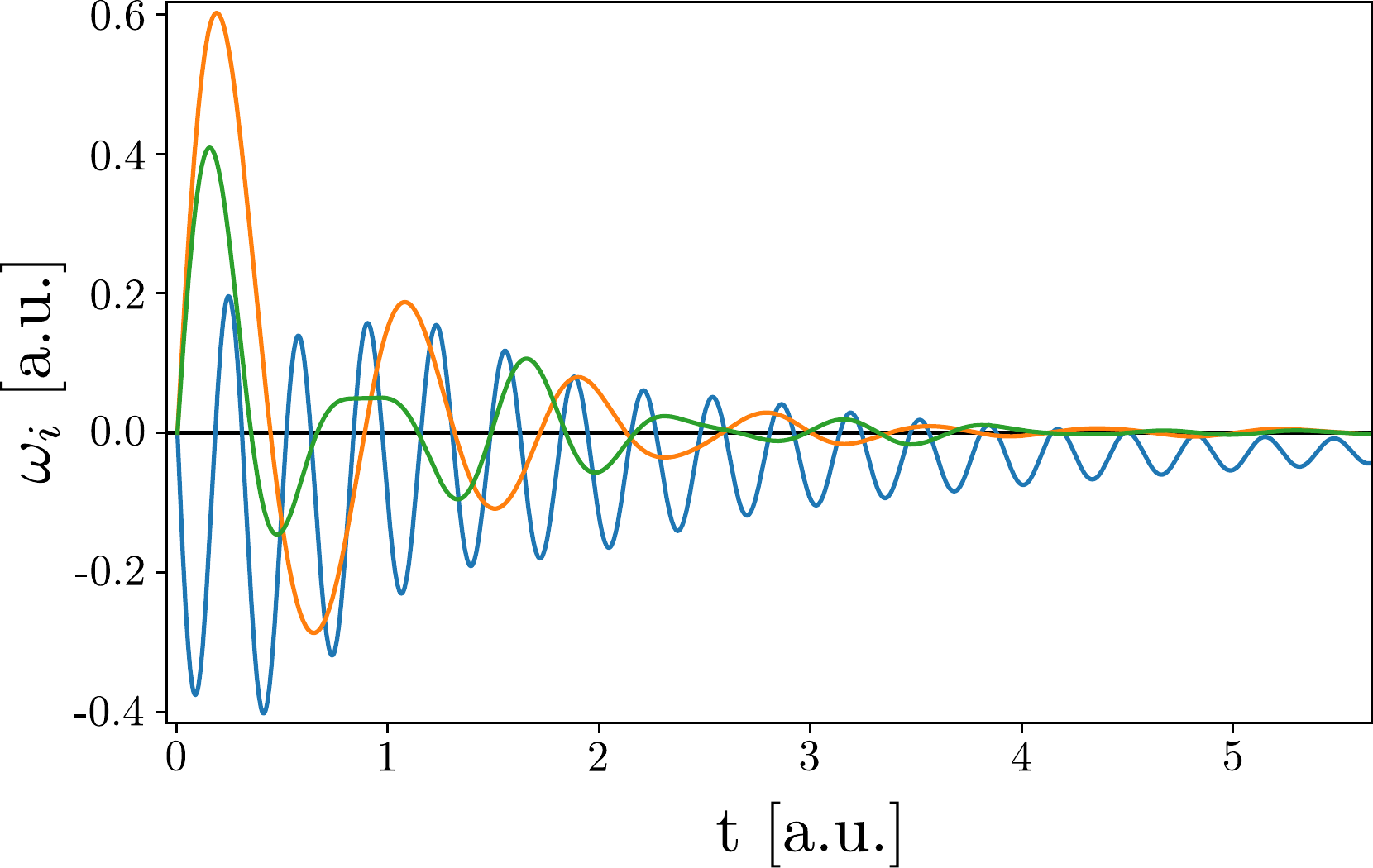}
 \caption{Time evolution of the frequency at the end node of the three lines whose removal produces the largest RoCoF in the IEEE 118-Bus test case. 
 For each line, the end node considered is the one with smallest inertia, leading to the maximal RoCoF according to Eq.~\eqref{eq:rocof_loc}. }
 \label{fig:freqs}
\end{figure}

In a damped system ($d_i\neq0$), the global RoCoF is maximal at time $t=0^+$, i.e., just following the disturbance.~\cite{Pag17} 
The same can be expected for the local RoCoF because, first, once the disturbance spreads through the network, its initial effect is diluted over more nodes, and second, the damping attenuates the amplitude of frequency oscillations, which generally implies a smaller derivative of the frequency. 
Our numerical data confirm that this is always the case (see Fig.~\ref{fig:freqs}), and the largest RoCoF is measured at the first time step following the line fault, in a $4^{\rm th}$-order Runge-Kutta implementation of Eq.~\eqref{eq:swing}. 

\begin{figure*}
 \centering
 \includegraphics[width=\textwidth]{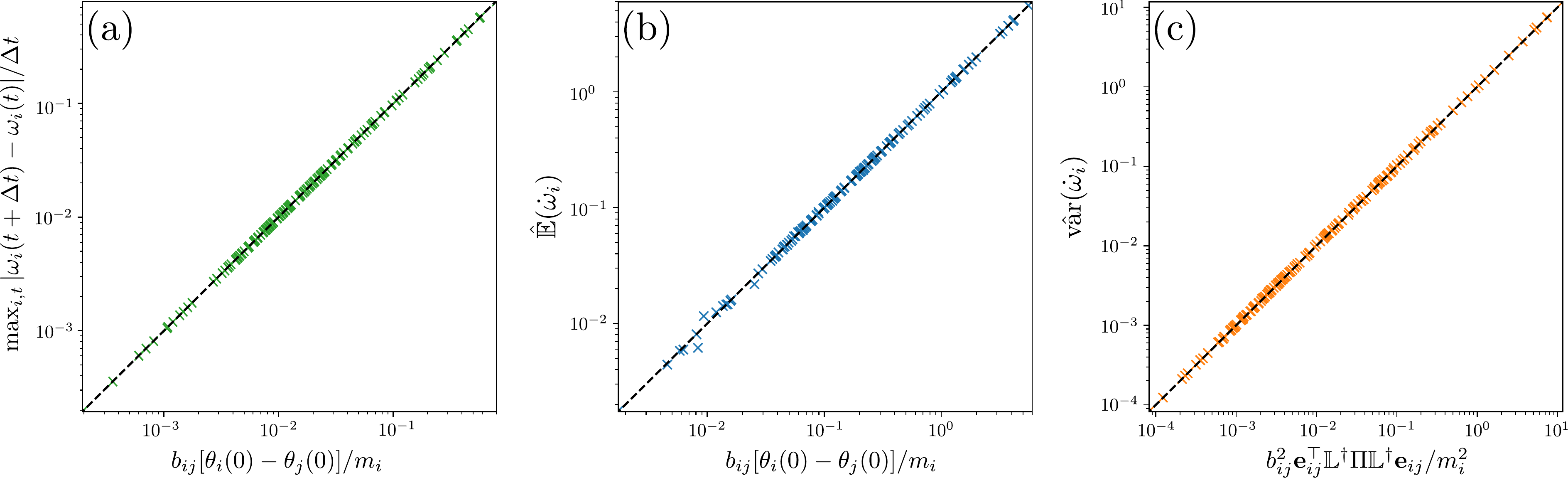}
 \caption{IEEE 118-Bus test case: 
 (a)~Numerically measured vs. analytically predicted RoCoFs after line contingencies. 
 Analytical predictions are given by Eq.~\eqref{eq:rocof_loc}. 
 (b)~Mean RoCoF over $10\,000$ random realizations of powers versus predicted expectations of the RoCoF, given by Eq.~\eqref{eq:E}. 
 (c)~Variance of the RoCoF over $10\,000$ random realizations of powers versus predicted variances of the RoCoF, given by Eq.~\eqref{eq:var}. 
 In each panel, each data point corresponds to the loss of one of the $170$ lines that do not fragment the network. }
 \label{fig:allin}
\end{figure*}

At $t=0^+$, the RoCoF can be computed straightforwardly, plugging initial conditions given by Eq.~\eqref{eq:init} in Eq.~\eqref{eq:swing}, and using Eq.~\eqref{eq:lap*}. 
We obtain 
\begin{align}\label{eq:rocof}
 M\dot{\bm{\omega}}(0) &= \bm{P} - \LL^*\LL^\dagger\bm{P} 
 = b_{ij}\bm{e}_{ij}\bm{e}_{ij}^\top\LL^\dagger\bm{P} \nonumber \\
 &= b_{ij}[\theta_i(0)-\theta_j(0)]\bm{e}_{ij}\, .
\end{align}
The only nonzero local RoCoFs just after the fault appear at the two ends of the lost line, and their value is the line load, divided by the inertia of the node, 
\begin{align}\label{eq:rocof_loc}
 \dot{\omega}_k &= (\delta_{ik}-\delta_{jk})\frac{b_{ij}[\theta_i(0)-\theta_j(0)]}{m_k}\, ,
\end{align}
where $\delta_{ab}$ is the Kronecker symbol.
Without loss of generality we take $m_i\leq m_j$, then node $i$ has the largest RoCoF. 
Based on the full knowledge of both the network and its initial operating state, Eq.~\eqref{eq:rocof_loc} allows one to identify the line which, when disconnected, leads to the largest RoCoF.
Fig.~\ref{fig:allin}(a) shows a perfect agreement between the prediction of Eq.~\eqref{eq:rocof_loc} and numerical results. 

\begin{figure*}
 \centering
 \includegraphics[width=\textwidth]{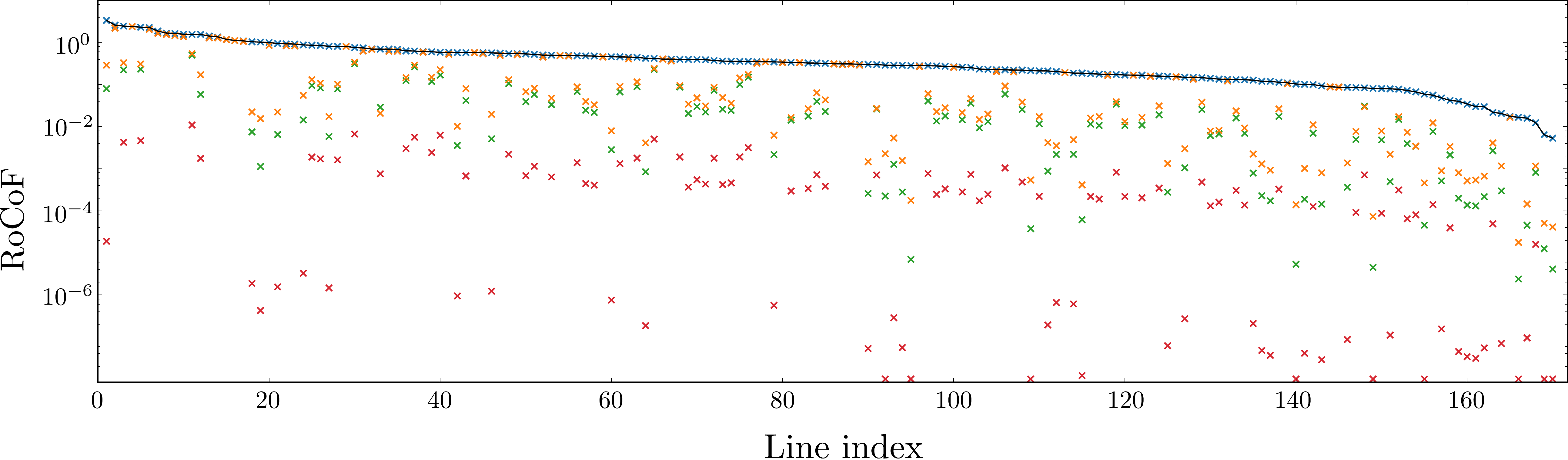}
 \caption{Maximal local RoCoF after the loss of each line that does not fragment the network in the IEEE 118-Bus test case. 
 Black line: theoretical RoCoF computed by Eq.~\eqref{eq:rocof}. 
 Blue crosses: maximal local RoCoF measured (simulation) among every nodes, with $100\%$ of the initial inertia at the loads. 
 Red (resp. green) crosses: maximal local RoCoF measured (simulation) among the generators, with $100\%$ (resp. $1\%$) of the initial inertia at the loads. 
 Orange crosses: maximal local RoCoF measured (simulation) among the generators, with inertia of the loads set to zero. } 
 \label{fig:reduced_inertia}
\end{figure*}

To confirm that our approach still gives reasonable predictions in the case where some nodes are inertialess, we numerically simulated the effect of line contingencies in networks where a set of loads have reduced or vanishing inertias. 
In this case, we measure the RoCoF only at nodes with finite inertia, because first, at inertialess nodes, the RoCoF is, in theory, infinite and thus makes no sense, and second, security measures based on RoCoF measurement are intended to protect rotating masses, i.e., with inertia, from mechanical damages. 
Results of the simulations for the IEEE 118-Bus test case are presented in Fig.~\ref{fig:reduced_inertia}. 
For each line contingency, we show, as a reference, RoCoF predicted by our theory (black line) and the maximal measured RoCoF (blue crosses) among every nodes, following the line fault, with initial inertia at every nodes. 
We then show the maximal measured RoCoF among the generators with the inertia of the loads being $100\%$ of their initial value (red crosses), $1\%$ of their initial values (green crosses), and zero (orange crosses). 
As can be expected, the lower the inertia at the loads, the larger the RoCoF at the generators. 
Nevertheless, we see that our prediction assuming inertia everywhere and computing the RoCoF at ever nodes is an upper bound for the maximal RoCoF. 
The most critical lines for the RoCoF are then the ones connected to at least one node with finite inertia. 

\subsection{Including uncertainties}
In the case of large transmission networks, the actual operating state cannot be exactly known. 
Therefore, we next assess the impact of a line loss, given some uncertainty on the distribution of power injections and consumptions. 
Our approach is statistical in that we never know $\bm{P}$ exactly, however we know the expectation value $\mu_k=\mathbb{E}(P_k)$ of each of its components, as well as the covariance matrix $\Pi_{k\ell}=\mathbb{E}\left[(P_k-\mu_k)(P_\ell-\mu_\ell)\right]$. 
This is all we need to know in order to derive the expectation and the variance of the RoCoF at node $i$ and $t=0^+$. We obtain
\begin{align}
 \mathbb{E}(\dot{\omega}_i) &= \frac{b_{ij}}{m_i}\bm{e}_{ij}^\top\LL^\dagger\bm{\mu}\, , \label{eq:E} \\
 {\rm var}(\dot{\omega}_i) &= \frac{b_{ij}^2}{m_i^2} {\rm var}(\bm{e}_{ij}^\top\LL^\dagger\bm{P}) 
 = \frac{b_{ij}^2}{m_i^2}\bm{e}_{ij}^\top\LL^\dagger\Pi\LL^\dagger\bm{e}_{ij} \nonumber \\ 
 &= \frac{b_{ij}^2}{m_i^2}\sum_{\alpha,\beta}\frac{(u_i^{(\alpha)}-u_j^{(\alpha)})(u_i^{(\beta)}-u_j^{(\beta)})}{\lambda_\alpha\lambda_\beta} {\bm{u}^{(\alpha)}}^\top \Pi \bm{u}^{(\beta)}\, . \label{eq:var}
\end{align}
Similar expressions with index permutations apply to the expectation and the variance of the RoCoF at node $j$. 
They vanish for all $k\notin\{i,j\}$, because in this work, we restrict ourselves to short-time calculation at $t=0^+$, before the perturbation propagates away from the faulted line. 

\begin{figure*}
 \centering
 \includegraphics[width=\textwidth]{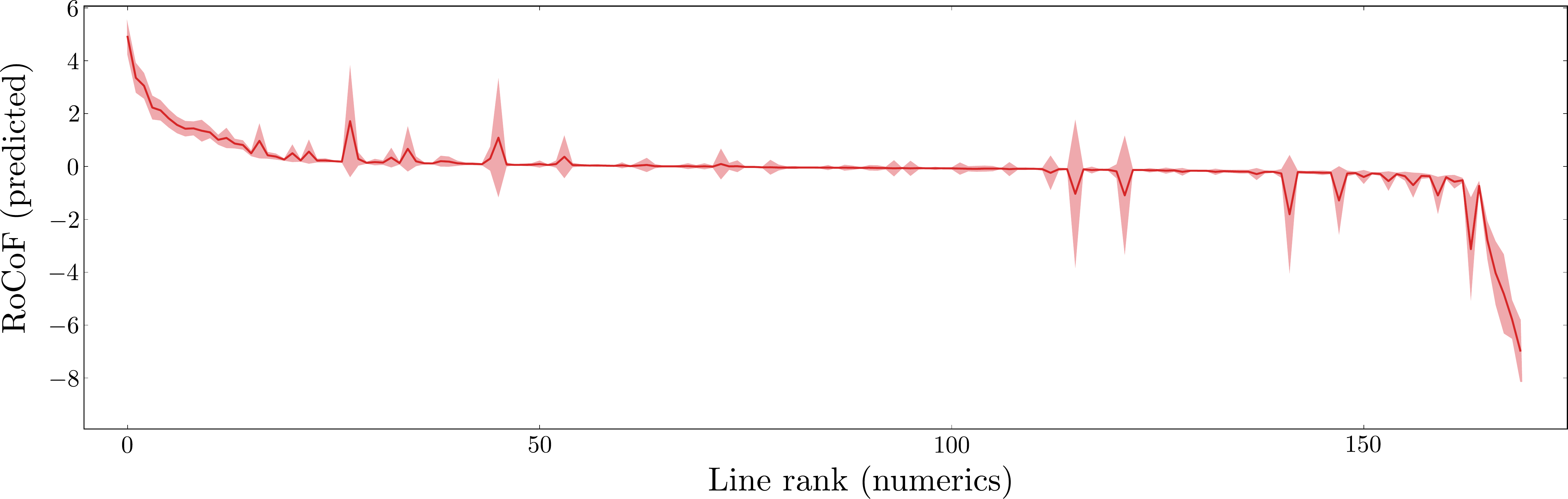}
 \caption{Theoretical mean RoCoF [Eq.~\eqref{eq:E}] as function of the numerical ranking of each line. 
 The shaded area represent $\mathbb{E}(\dot{\omega}_i) \pm \sigma_i$, with standard deviation $\sigma_i$ obtained by Eq.~\eqref{eq:var}. 
 The lines' ranking is obtained numerically. 
 For $10\,000$ power realizations, we rank the lines from largest to smallest RoCoFs following their loss. 
 We then compute a score for each line, which is the sum of its ranks among the $10\,000$ rankings obtained for the power realizations. 
 Finally, the lines are ranked from smallest to largest score. 
 Extreme scores mean that the lines are critical in most of the power profiles. }
 \label{fig:rocof_vs_score}
\end{figure*}

To confirm Eqs.~\eqref{eq:E} and \eqref{eq:var}, we took $10\,000$ random realizations of power profile $\bm{P}$, with uncorrelated components having a normal distribution with centroid $\mu_i=P_i$ and root mean square $\sigma_i=P_i/3$, corresponding to relatively large uncertainties. 
For each realization of $\bm{P}$, we simulated Eq.~\eqref{eq:swing} for each possible line loss and measured the maximal local RoCoF. 
Fig.~\ref{fig:allin}(b) shows the mean RoCoF calculated over the $10\,000$ power profile realizations, for each line loss, with respect to the expected RoCoF predicted by Eq.~\eqref{eq:E}, with $\bm{\mu}=\bm{P}$. 
Fig~\ref{fig:allin}(c) shows the variance of the RoCoF calculated over the $10\,000$ power profile realizations, for each line loss, with respect to the variance predicted by Eq.~\eqref{eq:var}.

We verified numerically the validity of our method to identify critical lines under uncertain power profile. 
Namely, for each realization of $\bm{P}$, we simulated the loss of each line and computed the maximal local RoCoF in each case. 
For each realization of $\bm{P}$, this gave us a ranking of the lines from the most critical (largest RoCoF) to the least critical (lowest RoCoF).
We then attributed a score to each line, computed as its mean rank over the $10\,000$ realizations. 
Fig.~\ref{fig:rocof_vs_score} shows the maximal theoretical mean [Eq.~\eqref{eq:E}] and the corresponding standard deviation [Eq.~\eqref{eq:var}] of the RoCoF following the loss of each line with respect to the ranking of the line according to their scores. 
The leftmost and rightmost lines are the ones whose loss provoked the largest RoCoF (in absolute value) in most of the simulations, and this was well predicted by the theoretical means and variances. 

\textbf{Remark.}
\textit{
It is interesting to note that in the special case where the covariance matrix can be decomposed as a Laurent polynomial in the Laplacian matrix, i.e., 
\begin{align}
 \Pi &= \sum_{k=-\infty}^{\infty}a_k\LL^k\, ,
\end{align}
the variance of the RoCoF [Eq.~\eqref{eq:var}] takes the elegant form 
\begin{align}
 {\rm var}(\dot{\omega}_i) &= \frac{b_{ij}^2}{m_i^2}\sum_{k=-\infty}^\infty a_k\Omega_{ij}^{(2-k)}\, ,
\end{align}
which we can write in terms of the resistance distances.~\cite{Kle93} 
The $q^{\rm th}$-order resistance distance between nodes $i$ and $j$ is defined as~\cite{Tyl18c}  
\begin{align}
 \Omega_{ij}^{(q)} &\coloneqq \bm{e}_{ij}^\top(\LL^\dagger)^q\bm{e}_{ij} & q &> 0\, , \nonumber \\
 &= \left[\left(\LL^\dagger\right)^q\right]_{ii}+\left[\left(\LL^\dagger\right)^q\right]_{jj} - 2\left[\left(\LL^\dagger\right)^q\right]_{ij}\, , \\
 \Omega_{ij}^{(q)} &\coloneqq \bm{e}_{ij}^\top\LL^q\bm{e}_{ij} & q &\leq 0 \nonumber \\
 &= \left(\LL^q\right)_{ii} + \left(\LL^q\right)_{jj} - 2\left(\LL^q\right)_{ij}\, .
\end{align}
This gives another example where performance measures in Laplacian-coupled dynamical 
systems depend on resistance distances and related quantities.~\cite{Tyl18a,Tyl18c}
}

\section{Conclusion}\label{sec:conclusion}
Considering the RoCoF as a measure of the impact of a perturbation on a network, we gave an analytical expression to assess the impact of a line contingency. 
The largest local RoCoF is observed at the time of the fault at an end of the lost line. 
It is proportional to the load of the line and inversely proportional to the inertia of the bus where it is measured [see Eq.~\eqref{eq:rocof_loc}]. 
Thus it directly depends on the current power profile of the system.  
As this power profile is generally unknown, we derived expressions for the expectation and the variance of the RoCoF, based on the expected power injections/consumptions and their covariance matrix. 

We overcame the challenge of an analytical treatment of line contingencies by relying on the initial graph Laplacian only. 
As our expressions, Eqs.~\eqref{eq:rocof_loc}, \eqref{eq:E}, and \eqref{eq:var}, depend on the eigenmodes of the network Laplacian matrix, only relying on the initial Laplacian significantly reduce the computation time, especially for large networks. 
Our results can then be used as a new tool to identify vulnerable lines under uncertain power profile. 

In the more general scope of coupled dynamical systems, we note that our method can be used to determine the initial response of a system, following the loss of a link. 
Namely, as soon as the dynamics of a network of coupled dynamical systems can be linearized with a Laplacian coupling, 
\begin{align}
 \frac{{\rm d}^q}{{\rm d}t^q}\bm{x} &= \sum_{k=1}^{q-1}a_k\frac{{\rm d}^k}{{\rm d}t^k}\bm{x} - \mathbb{L}\bm{x} + \bm{c}\, ,
\end{align}
the initial $q^{\rm th}$-order response can be obtained by mean of our method.

However, our results are limited to the time of the contingency ($t=0^+$). 
Even if the largest RoCoF occurs at this time at one end of the lost line, we cannot tell anything about the RoCoF at other nodes of the network. 
Weakening the assumptions made in our model would also give valuable information, in particular, including losses will change the dynamics and make it more realistic.
Fig.~\ref{fig:reduced_inertia} suggests that inertialess nodes do not have a jeopardizing effect on the RoCoF. 
However, an analytical approach including inertialess nodes is still lacking. 
Further research should focus on these aspects. 

This work was supported by the Swiss National Science Foundation under grants PYAPP2\_154275
and 200020\_182050.

\end{document}